\documentclass[twocolumn]{aastex702}

\usepackage{amsmath,amsfonts,amssymb}
\usepackage{booktabs}
\usepackage{soul} 
\usepackage{xcolor} 
\usepackage{hyperref}

\shorttitle{Kinetic Scaling of Magnetic Inverse Transfer}
\shortauthors{Cai et al.}

\begin{document}

\title{Inverse Transfer in Non-helical 2D Collisionless Magnetic Turbulence: Island-Merger Picture with Kinetic Effects}

\author[0000-0002-9074-4657]{Yangyang Cai}
\affiliation{Tsung-Dao Lee Institute, Shanghai Jiao Tong University, Shanghai 201210, China}
\email[show]{yangyang.cai@sjtu.edu.cn}

\author[0000-0002-2991-5306]{Hongzhe Zhou}
\affiliation{School of Mathematics, Physics and Statistics, Shanghai Polytechnic University, 2360 Jinhai Road, Shanghai, 201209, China}
\affiliation{Tsung-Dao Lee Institute, Shanghai Jiao Tong University, Shanghai 201210, China}
\email{hzzhou@sspu.edu.cn}

\author[0000-0002-8131-6730]{Yosuke Mizuno}
\affiliation{Tsung-Dao Lee Institute, Shanghai Jiao Tong University, Shanghai 201210, China}
\affiliation{School of Physics and Astronomy, Shanghai Jiao Tong University, Shanghai 200240, China}
\affiliation{Key Laboratory for Particle Physics, Astrophysics and Cosmology (MOE), Shanghai Key Laboratory for Particle Physics and Cosmology, Shanghai Jiao Tong University, Shanghai 200240, China}
\email[show]{mizuno@sjtu.edu.cn}

\correspondingauthor{Yangyang Cai, Yosuke Mizuno}

\begin{abstract}
Magnetic inverse transfer is often invoked to connect small-scale magnetic-field generation to larger coherence scales in high-energy and cosmological plasmas.
The underlying magnetohydrodynamic (MHD) arguments combine two logically distinct ingredients:
a bulk quantity that is asymptotically conserved in the limit of small resistivity,
and a time scale determined by the decay dynamics.
In this work, we explore whether this scenario still holds in decaying nonhelical turbulence formed by collisionless plasmas using particle-in-cell simulations.
The simulations approximately satisfy $B^2\xi_B^2\simeq{\rm const}$ as in the MHD case, and the fitted exponents in $B^2\propto t^{-p}$ and $\xi_B\propto t^q$ obey $p\simeq2q$.
Here $B^2\equiv\langle B_x^2+B_y^2\rangle$ is the average in-plane magnetic energy density, and $\xi_b$ is the magnetic integral scale.
However, the decay time scale differs from the MHD case as inferred from the decay exponents.
We found $p<1$ and $q<1/2$ in all cases with different initial magnetization $\sigma_0$, with both exponents lower than the MHD values and varying systematically with $\sigma_0$.
The spectral peak also migrates toward lower wavenumber at a rate faster than the growth of $\xi_B$, indicating a broken self-similarity. The broken self-similarity is attributed to the appearance of kinetic scales in the magnetic energy spectrum due to pressure anisotropy and Larmor-scale magnetic structures. These results indicate that in astrophysical collisionless plasmas, including but not restrict to solar wind, pulsar-wind nebulae, interstellar medium, and cosmological plasmas, magnetic coherence can continue to grow by inverse transfer, but extrapolations based on MHD decay-time scaling can overestimate the rate of large-scale field growth.
\end{abstract}

\section{Introduction}

Freely decaying magnetic turbulence can transfer magnetic energy from the energy injection scale toward larger spatial scales. In the early Universe, inverse transfer changes the survival and large-scale imprint of primordial magnetic fields \citep{banerjee2004,brandenburg2015,brandenburg2017}. In high-energy astrophysical plasmas, the growth of magnetic coherence scales affects how magnetic energy is stored, dissipated, and converted into radiation or nonthermal particles in systems such as gamma-ray bursts, blazars, relativistic jets, and pulsar-wind nebulae \citep{zrake2014,bhat2021}. These applications motivate the same theoretical question: whether a magnetic field generated on small or intermediate plasma scales can move to larger scales, and, if they do, what are the laws that govern this process.

In the framework of magnetohydrodynamics (MHD), inverse transfer in decaying magnetic turbulence has been studied from several related viewpoints. Early work used scaling and self-similarity arguments to connect the initial magnetic spectrum to the subsequent growth of the correlation length \citep{olesen1997,son1999}. Direct simulations of decaying MHD turbulence then established self-similar decay and inverse transfer, especially for helical fields \citep{christensson2001,banerjee2004}. More recent work showed that even non-helical magnetic turbulence can exhibit inverse transfer \citep{brandenburg2015,zrake2014,brandenburg2017}. \citet{hosking2021} showed that in reconnection-controlled non-helical decay, a helicity-fluctuation integral (Hosking integral) can act as an approximate invariant and, when combined with self-similarity, determines the decay exponents. \citet{zhou2022hosking} then tested this quantity directly in decaying magnetically dominated turbulence, confirming its gauge invariance and approximate conservation at large Lundquist numbers. They also showed that the measured decay exponents evolve near the self-similar line expected from the Hosking integral. In this invariant-based viewpoint, the large-scale evolution is fixed by a conserved quantity plus a self-similar spectral form.

A complementary viewpoint from local dynamics emphasizes magnetic-island coalescence. \citet{zhou2019} proposed that in 2D MHD, inverse transfer can be understood as a hierarchy of magnetic-island mergers controlled by reconnection. We denote by $B$ the root-mean-square (rms) in-plane magnetic field and by $\xi_b$ the magnetic integral scale measured from the in-plane magnetic-energy spectrum.
In the simplest picture, with initial magnetic islands having the same size and flux, $\xi_b$ is proportional to the island radius $R_n$ in merger generation $n$, and $B$ is proportional to the reconnecting field $B_n$. When two equal islands merge, assuming incompressible fluid, the area doubles, thus
\begin{equation}
  R_{n+1}=\sqrt{2}\,R_n .
\end{equation}
Conservation of island flux(2D flux enclosed in an island) gives $B_nR_n\simeq{\rm const.}$, hence
\begin{equation}
  B_{n+1}=B_n/\sqrt{2}.
\end{equation}
The local merger time is estimated from a reconnection time,
\begin{equation}
  \tau_n\sim \frac{R_n}{\epsilon_{\rm rec}v_{A,n}},
\end{equation}
where $\epsilon_{\rm rec}$ is the normalized reconnection rate. In the idealized MHD hierarchy, $\epsilon_{\rm rec}$ is taken to be generation independent and, for fixed density and no additional scales, $v_{A,n}\propto B_n$. Therefore
\begin{equation}
  \frac{\tau_{n+1}}{\tau_n}
  =
  \frac{R_{n+1}/v_{A,n+1}}{R_n/v_{A,n}}
  =2 .
\end{equation}
Since $\xi_b$ grows as the island radius and $B^2$ decreases as the square of the reconnecting field, after $n$ merger generations, one has $\xi_b\propto2^{n/2}$ and $B^2\propto2^{-n}$. The relation $\tau_{n+1}/\tau_n=2$ gives $t\propto2^n$, so the corresponding continuous scalings are
\begin{equation}
  \xi_b\propto t^{1/2},\qquad B^2\propto t^{-1}.
\end{equation}

\citet{zhou2021} generalized the island-merger picture by writing a kinetic equation for the statistical distribution $f(A,\psi,t)$ of islands with area $A$ and flux $\psi$. In that formulation, the growth of the mean island area, the decay of magnetic energy, and the migration of the magnetic scale are moments of the full island distribution. In this sense, inverse transfer is a population-level process: a local X-point reconnection rate controls an individual merger, but the global decay rate also depends on how many islands of different sizes are available to merge, how their fluxes are distributed, and what fraction of local reconnection events contributes to coherent large-scale coalescence.

The invariant-based and island-merger descriptions therefore share a common structure. First, a conserved or slowly varying quantity fixes a power-law relation between energy and scale. Second, self-similarity supplies a time scale: the normalized spectrum or island distribution is a rescaled copy of itself, the typical reconnecting structure is proportional to the magnetic integral scale, and, most importantly, the global inverse-transfer time scale is uniquely determined by the a local merger or reconnection time scale.

Dynamics in collisionless kinetic turbulence is far richer, and the single time scale evolution, as seen in decaying MHD turbulence, can be easily broken.
Collisionless conditions are common in high-energy dilute plasmas, where particle mean free paths can exceed the macroscopic magnetic-correlation scale and the magnetic field decay is controlled by kinetic effects rather than resistivity. Such plasmas contain pressure anisotropy, finite Larmor radius effects, and skin-depth-scale current sheets. Pressure anisotropy can reduce the effective magnetic tension through invariants proposed by Chew--Goldberger--Low (CGL) model\citep{cgl1956} and can drive firehose or mirror instabilities near kinetic scales \citep{schekochihin2008,kunz2014,melville2016,squire2017}. Kinetic transition-range spectra are also known to steepen near ion-Larmor or sub-ion scales in theoretical, observational, and particle-in-cell (PIC) studies \citep{voitenko2011,sahraoui2010,sahraoui2009,zhdankin2017,zhdankin2018}. Recent work has further suggested that pressure-anisotropy-driven instabilities can suppress inverse magnetic-energy transfer in marginally magnetized collisionless plasmas by reducing magnetic tension and inhibiting reconnection-driven coalescence \citep{zhou2025}.

In this paper, we perform 2.5-dimensional (2.5D) PIC simulations of decaying pair-plasma turbulence. We try to address four questions, following the order of the analysis:
(i) what conservation law governs the decay?
(ii) is the magnetic spectrum compatible with a single self-similar scale?
(iii) do the measured decay exponents follow the equal-island MHD predictions? (iv) if the exponents depart from the MHD prediction, what causes the difference?

The rest of the paper is organized as follows. Section 2 describes the PIC simulation setup. Section 3 tests the conservation constraint. Section 4 measures the decay exponents. Section 5 analyzes spectrum indices and the degree of self-similarity. Section 6 discusses two factors that affect the decay exponents. Section 7 discusses the plasma-astrophysical implications and limitations, and Section 8 gives the conclusions. Unless stated otherwise, all quantities are reported in Gaussian units with $c=1$.

\section{Numerical Setup}

We perform 2.5D PIC simulations using the GPU-accelerated relativistic PIC code \texttt{Entity} \citep{hakobyan2025entity}.
The simulation domain is a periodic square in the $x$--$y$ plane of side length $L$ with a uniform guide field $B_\text{g}\hat{\bm z}$. All three components of the electromagnetic fields and particle velocities are evolved, while all quantities are independent of $z$. All field components are normalized by $B_\text{g}$, and we will use the lowercase letter $b\equiv {B}/{B_\text{g}}$ to represent normalized magnetic fields. We always set the initial root-mean-square (rms) in-plane magnetic field as $b_{\rm rms,0}=B_{\rm rms,0}/B_\text{g}=0.5$. For simplicity, we use electron-positron pair plasmas with initial temperature $k_B T/mc^2=0.3$. The grid resolution is $8192^2$, and the cold skin depth $d_0$ satisfies
\begin{equation}
  \frac{L}{d_0}=5120.
\end{equation}
The grid resolves this scale with approximately $1.6$ cells per $d_0$.

The initial in-plane magnetic fluctuations follow a modified version of the setup used by \citet{comisso2018}. We construct a divergence-free random field from a periodic flux function,
\begin{equation}
  A_z(x,y)
  =
  \sum_{i,j=0}^{N_2-N_1}
  a_{ij}
  \cos(k_i x+\phi_{ij})
  \cos(k_j y+\varphi_{ij}),
\end{equation}
where
\begin{equation}
  k_i=\frac{2\pi(N_1+i)}{L},
  \,
  k_j=\frac{2\pi(N_1+j)}{L},
\end{equation}
and
\begin{equation}
  a_{ij}=\frac{\alpha}{(k_i^2+k_j^2)^{1/2}} .
\end{equation}
The phases $\phi_{ij}$ and $\varphi_{ij}$ are randomly chosen from $[0,2\pi)$. The initial in-plane field is
\begin{equation}
  b_x=\frac{\partial A_z}{\partial y},
  \qquad
  b_y=-\frac{\partial A_z}{\partial x},
\end{equation}
so that $\nabla\cdot\boldsymbol{b}=0$ by construction. The amplitude is chosen as
\begin{equation}
  \alpha=\frac{2b_{\rm rms,0}}{M},
  \qquad
  M=N_2-N_1+1,
\end{equation}
which gives $b_{\rm rms,0} = \langle b_x^2+b_y^2\rangle^{1/2}$ on the periodic grid at $t=0$. Here $\langle...\rangle$ denotes a spatial average over the $x-y$ plane, as in the following of this paper.
The excited mode band is between
\begin{equation}
  N_1=33,\, N_2=64,
\end{equation}
and we define the reference length
\begin{equation}
  l_0 = L/N_2.
\end{equation}

\begin{figure*}
  \centering
  \includegraphics[width=0.98\textwidth]{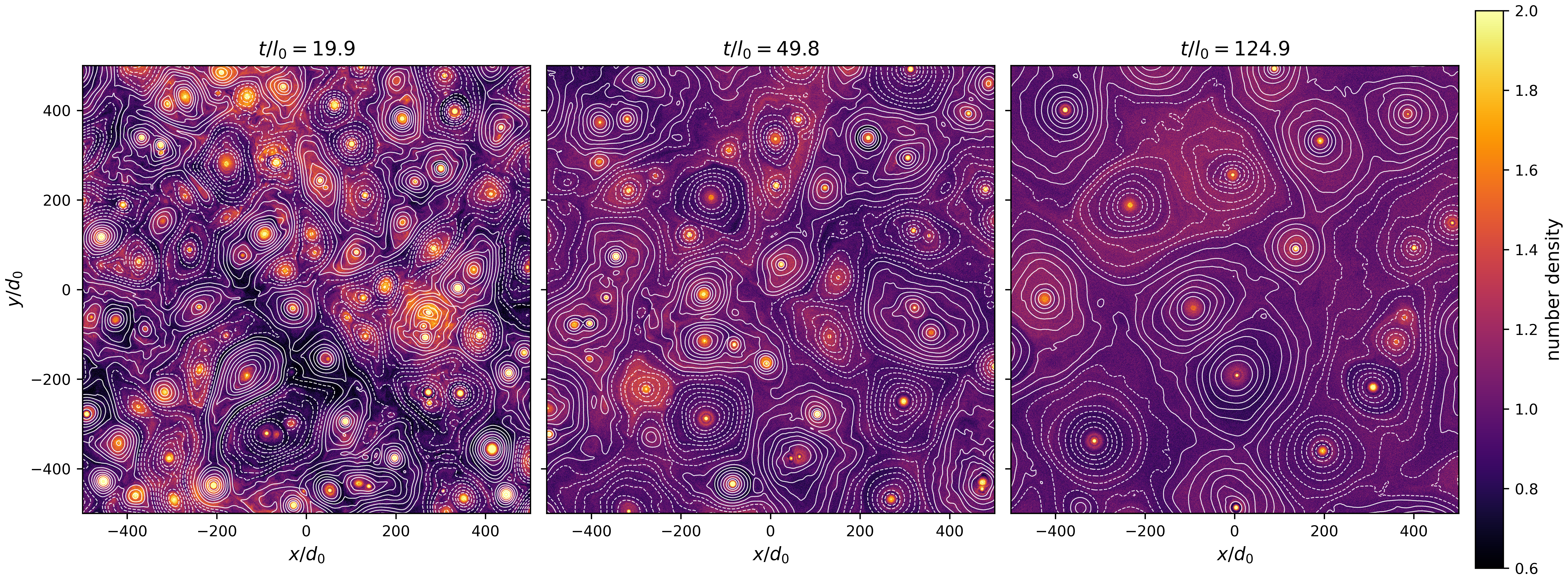}
  \caption{Evolution of a representative $\sigma_0=1$ simulation. For a clear view of the islands, only a $(500d_0)^2$(Full length $L=5120d_0$) area is presented. Colors show
  the total number density on a linear scale, and
  white contours indicate the in-plane magnetic flux function $A_z$, reconstructed
  from $b_x=\partial_y A_z$ and $b_y=-\partial_x A_z$. The solid and dashed lines show positive and negative $A_z$, respectively. The three snapshots,
  at $t/l_0=19.9$, $49.8$, and $124.9$, present the coalescence of small magnetic
  islands into fewer and larger structures.}
  \label{fig:island_evolution}
\end{figure*}

We consider five initial values of background field magnetization,
\begin{equation}
  \sigma_0 = 0.25,\ 1,\ 4,\ 16,\ 64.
\end{equation}
Here $\sigma_0=B_\text{g}^2/4\pi n_0$, where $n_0$ denotes the initial number density, is the magnetization corresponding to the uniform guide field. The production runs use 128 particles per cell. For the $\sigma_0=16$ case, we also performed convergence tests with grid resolution as high as $16392^2$ and particle number as high as 256 particles per cell. These higher-resolution and higher-particle-number tests reduce particle noise, but do not change the physical conclusions reported below.
Figure~\ref{fig:island_evolution} illustrates the evolution of a representative
$\sigma_0=1$ simulation through the main dynamical stage. The initially
numerous magnetic-flux structures coalesce into fewer and larger islands. 

\section{Conservation Constraint}

We first look into whether a conserved quantity exists in the decaying kinetic turbulence.
For this purpose, we consider the evolution of the in-plane magnetic energy
\begin{equation}
  b^2=\langle b_x^2+b_y^2\rangle,
\end{equation}
and the magnetic integral scale defined from the in-plane magnetic energy spectrum $E_b(k)$,
\begin{equation}
  \xi_b =
  \frac{\int E_b(k)k^{-1}\,\text{d}k}
       {\int E_b(k)\,\text{d}k}.
\end{equation}
The same spectrum is used for the peak and integral-scale diagnostics below.

\begin{figure}
  \centering
  \includegraphics[width=1.0\linewidth]  {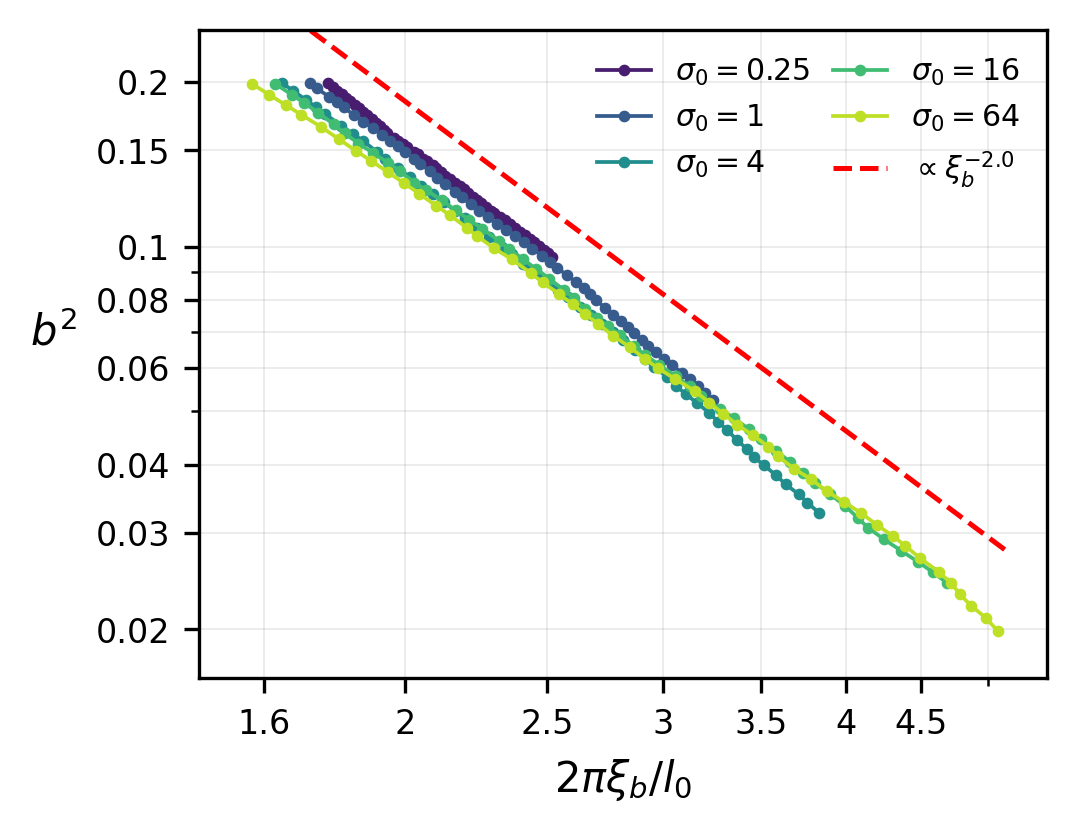}
  \caption{Relation between the in-plane magnetic-energy proxy $b^2$ and the magnetic integral scale $\xi_b$ for all five magnetizations, starting each case at the first sample with $b^2\simeq0.2$. For reference, the red dashed line has slope $-2.0$. The approximate alignment with this slope indicates $b^2\xi_b^2\simeq{\rm const}$.}
  \label{fig:bxy_xi}
\end{figure}

In 2D MHD island-merger models, the approximate conservation of island flux leads to the scaling, 
$b^2\xi_b^2\simeq{\rm const}$, during the merger hierarchy \citep{zhou2019,zhou2021}. More generally, invariant-based descriptions of decaying MHD turbulence also emphasize that inverse transfer is constrained by slowly varying magnetic invariants, although the relevant invariant need not be the same in different systems \citep{hosking2021,zhou2022hosking}. For the collisionless plasmas considered in this work, when plotting $b^2$ against $\xi_b$, Figure~\ref{fig:bxy_xi} shows that
$
  b^2 \propto \xi_b^{-\alpha}
$
where $\alpha \approx 2$. Equivalently,
\begin{equation}
  b^2\xi_b^2\simeq {\rm const},
  \label{eq:flux_constraint}
\end{equation}
in consistency with the energy--scale relation in the island-merger picture: larger magnetic islands contain weaker reconnecting fields while the product $b\xi_b$ remains approximately constrained.

The agreement with the MHD scaling should not be over-interpreted as proof of an identical microscopic invariant in collisionless plasmas.
In the theory of MHD, the conserved flux is subject to incompressibility (i.e., island mergers conserve mass), which is not generally true in the collisionless plasmas. We therefore use Equation~(\ref{eq:flux_constraint}) as an empirical conservation-like constraint on the bulk dynamics, not as a claim that the collisionless system conserves precisely the same MHD quantity.

\section{Decay Exponents}

The previous section shows that a conservation law exists to constrain magnetic energy and length scale evolution in Equation~(\ref{eq:flux_constraint}). That constraint, however, does not determine the decay exponents $p$ or $q$ in the power law evolution,
\begin{align}
    b^2\sim t^{-p},\qquad \xi_b\sim t^q,
\end{align} 
but only a combined relation $p\simeq2q$. In MHD models, it only follows from self-similarity that the global decay time is proportional to the local merger time, giving $q=1/2$ and $p=1$
\citep{zhou2019}.

\begin{figure*}
  \centering
\includegraphics[width=1.0\textwidth]{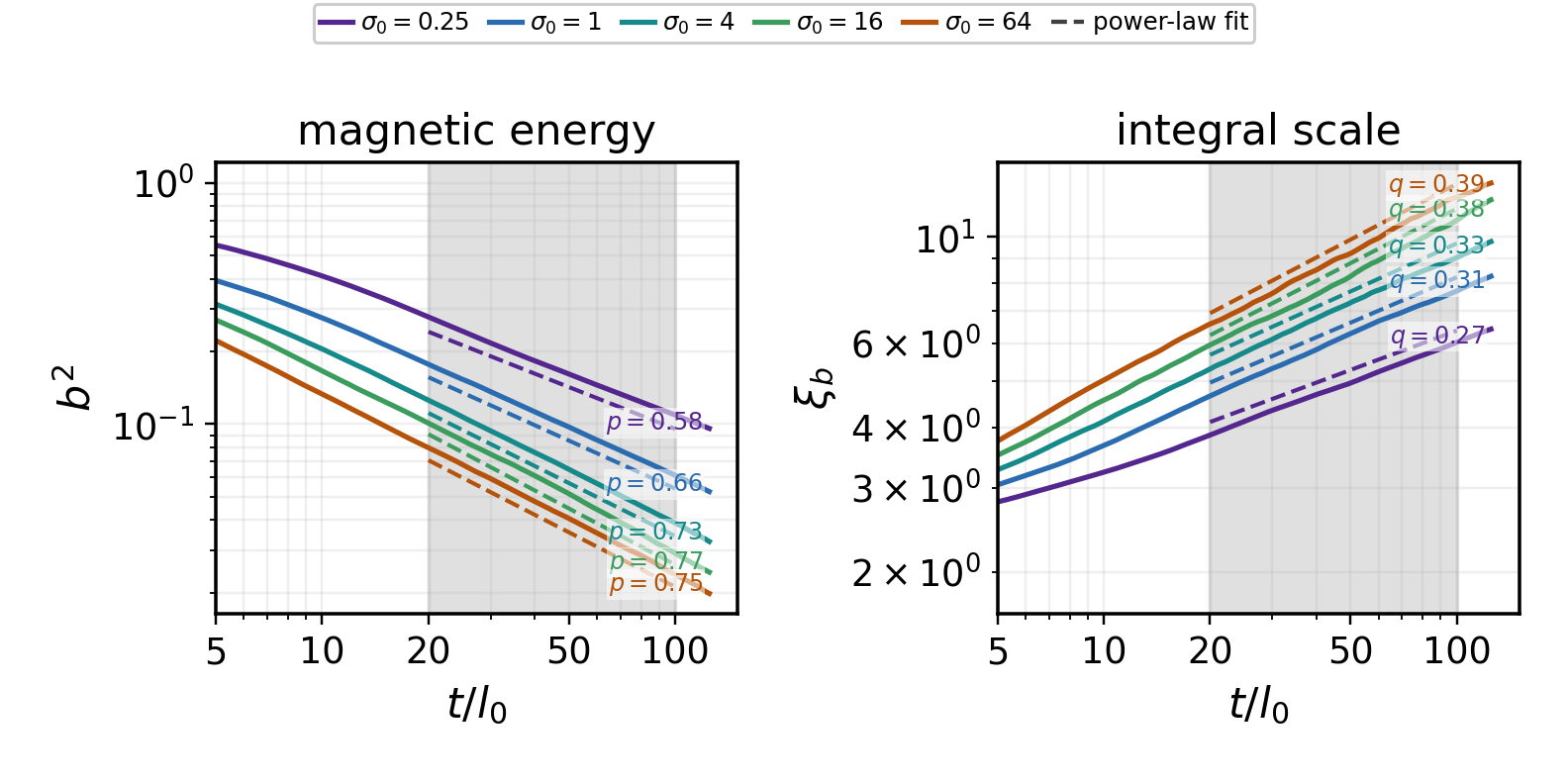}
  \caption{Power-law fits for $b^2(t)$ and $\xi_b(t)$. Colored solid curves show the simulation data, same-color dashed lines show the power-law fits, and the shaded region marks the fitted interval $20\leq t/l_0\leq100$. Text labels give the fitted exponents.}
  \label{fig:powerlaw_fits}
\end{figure*}

Figure~\ref{fig:powerlaw_fits} shows that both $b^2$ and $\xi_b$ evolve as power laws. The fitted curves track both $b^2(t)$ and $\xi_b(t)$ closely inside the shaded interval. Thus, a single power law is an adequate description of the main decay stage. Figure~\ref{fig:powerlaw_fits} also shows the results of the linear fitted power-law exponents. All cases have $p<1$ and $q<1/2$, indicating that the inverse transfer is slower than the MHD prediction. Interestingly, the power-law exponents are different for different $\sigma_0$, which was not predicted by any previous MHD theories.

\begin{figure}
  \centering
  \includegraphics[width=1.0\linewidth]{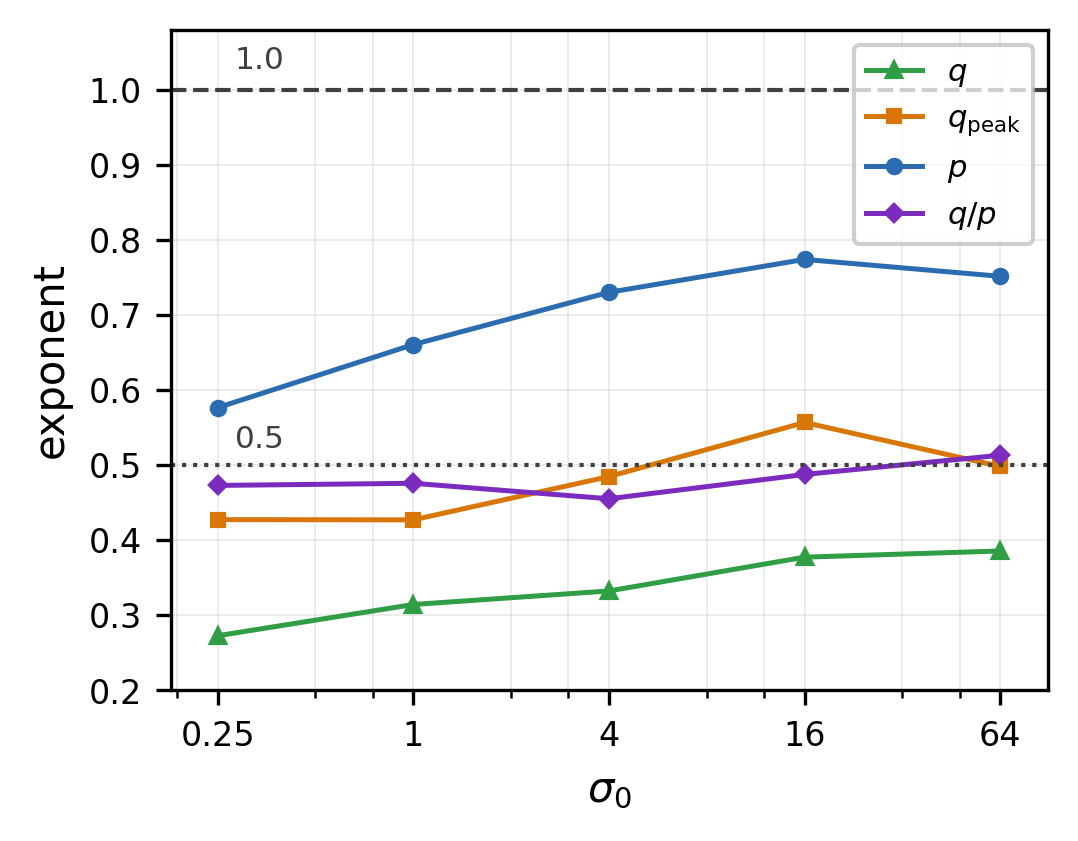}
  \caption{The exponents shown in this figure are defined by $b^2\sim t^{-p}$, $\xi_b \sim t^q$, and $k_{\rm peak}\sim t^{-q_{\rm peak}}$. The purple line represents the ratio $q/p$, confirming the conservation law. The black dotted and black dashed lines mark 0.5 and 1.0, respectively, for reference.}
  \label{fig:decay_exponents}
\end{figure}

Figure~\ref{fig:decay_exponents} plotted the exponents versus $\sigma_0$, and also the ratio of $q/p$. The value of $q/p$ is approximately 0.5 and does not show any significant dependence on $\sigma_0$, which is consistent with the conservation law Equation\eqref{eq:flux_constraint} analyzed in the previous section.
Both $p$ and $q$ show strong dependence with $\sigma_0$ and seem to reach a limit at high $\sigma_0$.

Furthermore, the motion of the spectral peak provides a useful reference of inverse transfer. We define $k_{\rm peak}$ as the wavenumber at which $E_b(k,t)$ is maximal, and fit for
\begin{equation}
  k_{\rm peak}\propto t^{-q_{\rm peak}}
\end{equation}
over the same main interval, $20\leq t/l_0\leq100$. Figure~\ref{fig:decay_exponents} shows that
\begin{equation}
  q_{\rm peak}\simeq
  0.43,\ 0.43,\ 0.48,\ 0.56,\ 0.50
\end{equation}
for $\sigma_0=0.25,\ 1,\ 4,\ 16,\ 64$, respectively. These values are closer to the MHD peak-scale expectation $q_{\rm peak}=1/2$ \citep[e.g.,][]{zhou2019}. In contrast, a similar fitting for the integral scale $\xi_b\propto t^{-q}$ is systematically lower than $1/2$,
and reaches $\sim0.4$ for high magnetization. It indicates that the decay is governed by more than one length scale and different scales evolve at different power-law rates.

\section{Spectrum and Self-Similarity}

\begin{figure*}
  \centering
  \includegraphics[width=0.75\linewidth]{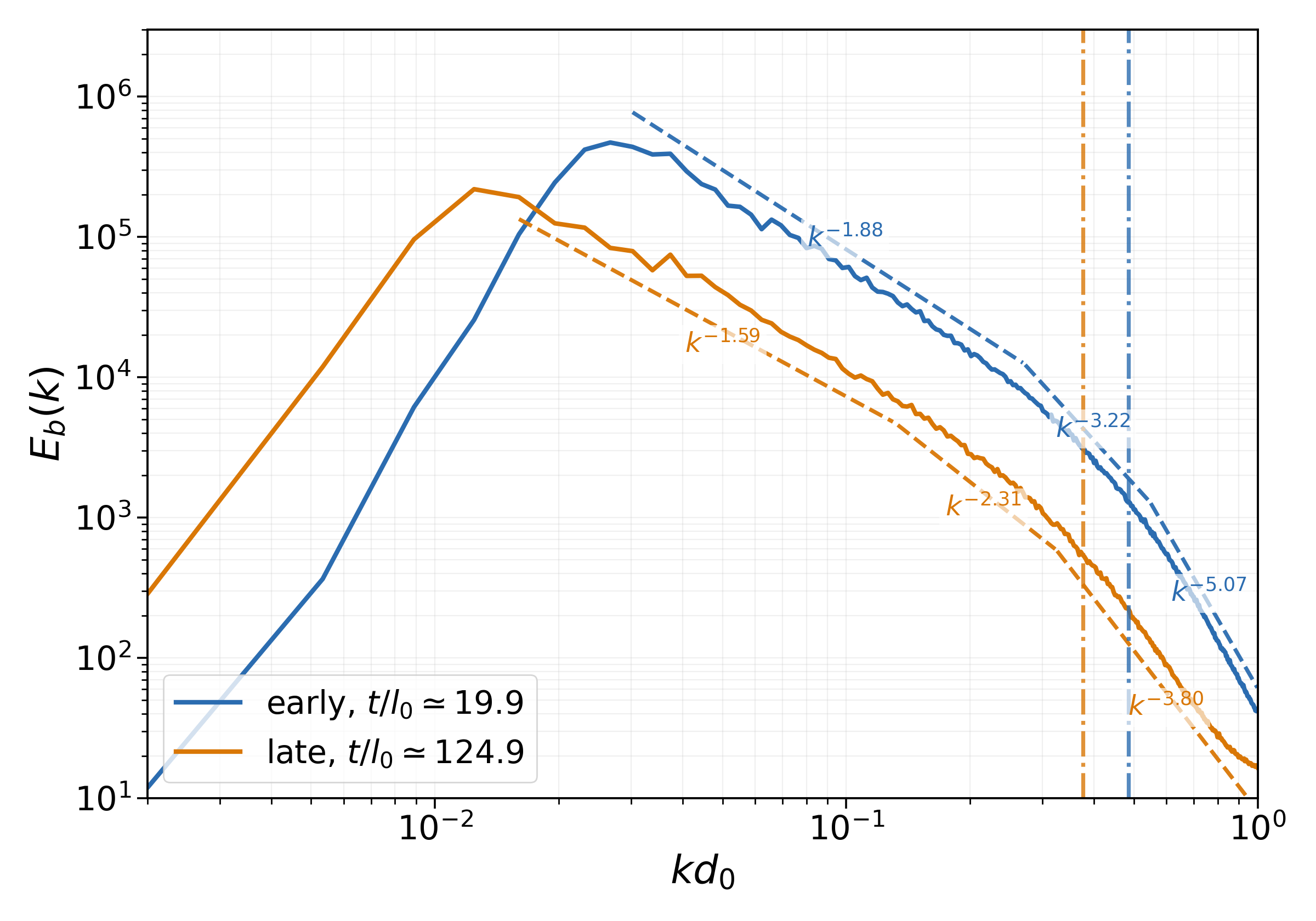}
  \caption{Representative magnetic-energy spectra for $\sigma_0=1$ at early ($t/l_0\simeq19.9$) and late ($t/l_0\simeq124.9$) times. Dashed lines show three-piece log-log fits to the post-peak spectrum over the resolved range $kd_0\leq1$. Dash-dotted vertical lines mark $k\rho_L=1$ for the two displayed times, where $\rho_L$ is the averaged Larmor radius of particles. The listed indices show that the post-peak spectral shape evolves during the decay.}
  \label{fig:full_spectrum}
\end{figure*}

We now turn to the spectral shape. Figure~\ref{fig:full_spectrum} shows two representative magnetic-energy spectra for $\sigma_0=1$, one near the beginning of the main dynamical interval ($t/l_0\simeq19.9$) and one near the end of the displayed interval ($t/l_0\simeq124.9$). The energy-containing peak moves toward smaller $k$, indicating inverse transfer. More importantly for the present argument, the fitted post-peak spectral indices (i.e., the spectral slopes to the right of the peak) change from approximately $(-1.88,-3.22,-5.07)$ at $t/l_0\simeq19.9$ to $(-1.59,-2.31,-3.80)$ at $t/l_0\simeq124.9$. A strictly self-similar spectrum would preserve its normalized shape and would not exhibit such a change of post-peak indices.

\begin{figure*}
  \centering
  \includegraphics[width=0.95\textwidth]{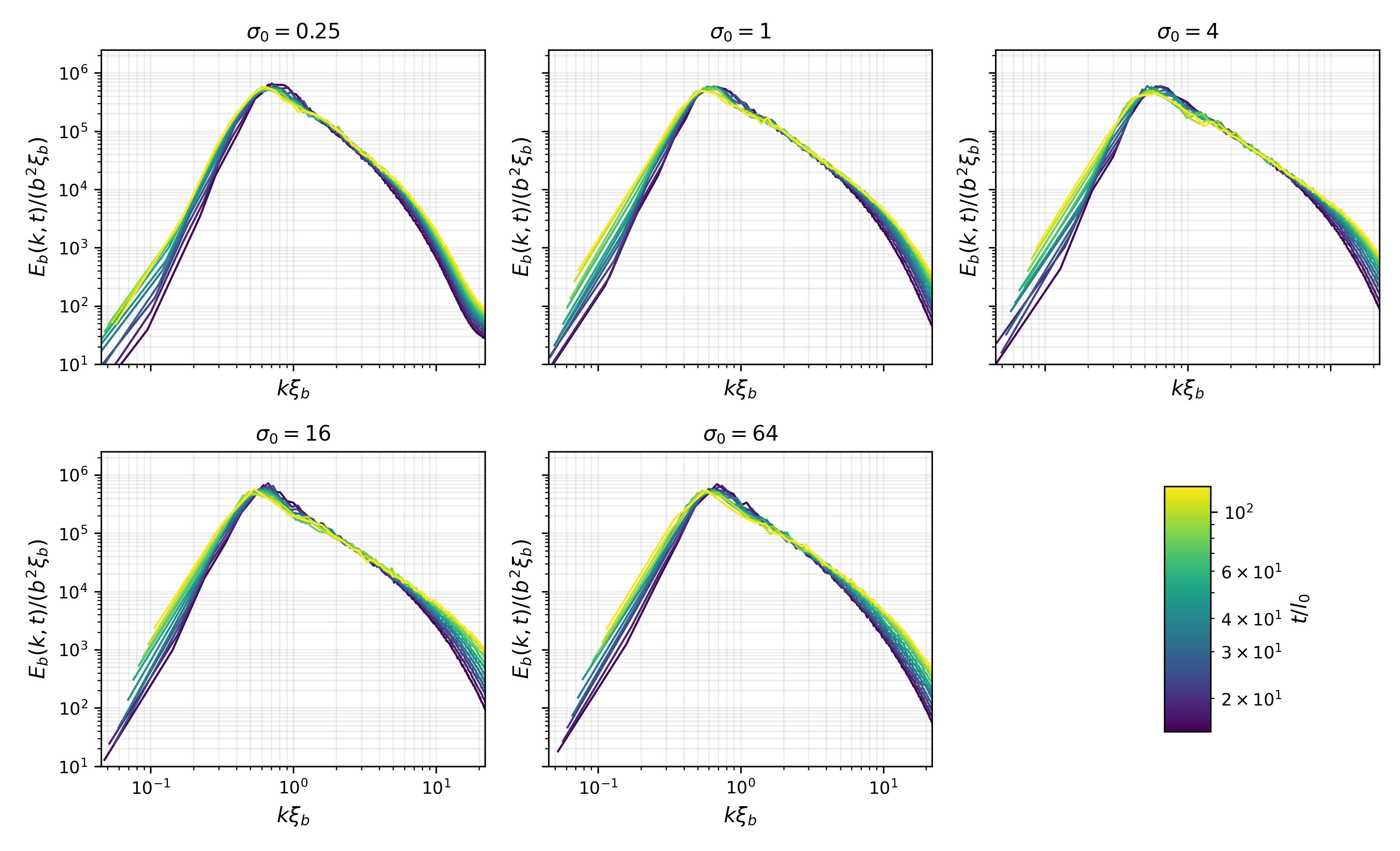}
  \caption{Self-similar spectrum-collapse test. Spectra are plotted as $E_b(k,t)/(b^2\xi_b)$ versus $k\xi_b$ for the main dynamical interval. The energy-containing range collapses well, while the low- and high-$k\xi_b$ tails retain systematic residuals.}
  \label{fig:spectrum_collapse}
\end{figure*}

A direct test on the spectral shape invariance further confirms the absence of self-similarity.
Strict single-scale decay would imply
\begin{equation}
  E_b(k,t)=b^2(t)\xi_b(t)\Phi(k\xi_b),
  \label{eq:single_scale_spectrum}
\end{equation}
where the factor $b^2\xi_b$ follows from $\int E_b\,dk\sim b^2$ and a spectral width of the order $\xi_b^{-1}$,
and $\Phi$ is a dimensionless shape function \citep{olesen1997}.
Figure~\ref{fig:spectrum_collapse} plots the dimensionless form $E_b/b^2\xi_b$ and tests whether the spectral shape $\Phi$ is indeed invariant. While the bulk, energy-containing spectrum is approximately self-similar, the full spectrum is not a single function of $k\xi_b$. In particular, the post-peak side retains a time-dependent shape, suggesting the existence of additional length scales.
We note that although the low-$k$ part is also time dependent, it is significantly affected by the finite box size and thus unphysical. Finite particle number noise becomes significant when $kd_0>1$; thus, we also ignored this part of the spectrum.

\begin{figure*}
  \centering
  \includegraphics[width=0.95\textwidth]{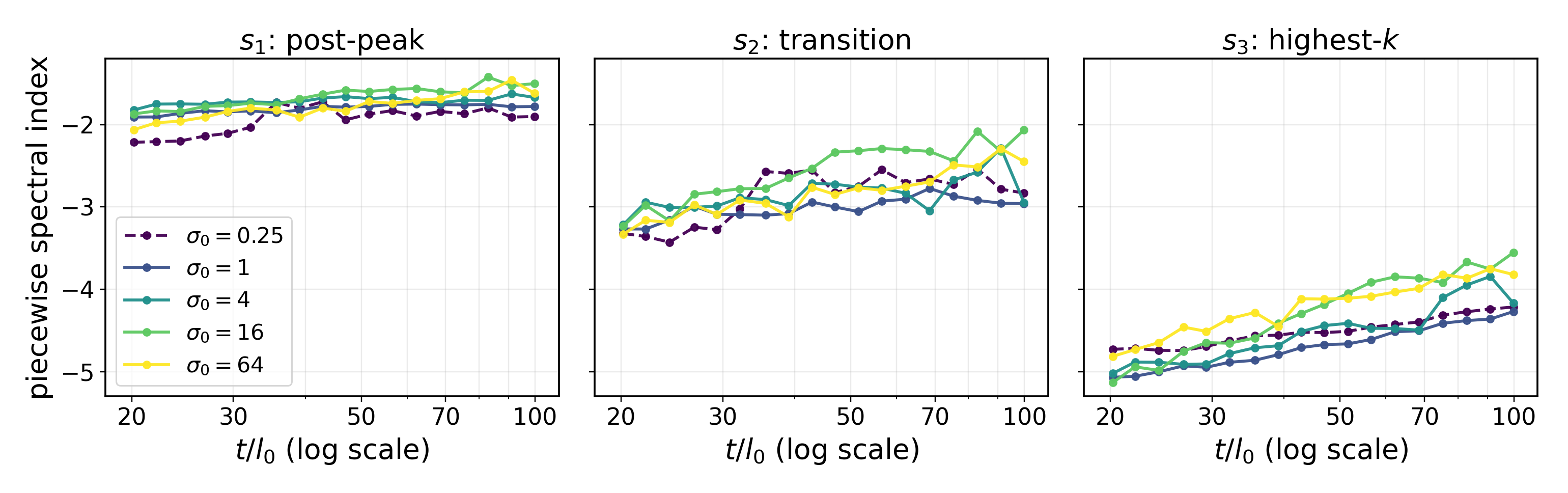}
  \caption{Log-time evolution of the fitted post-peak spectral indices. The fit uses $kd_0\leq1$. The three panels show the lower-$k$ post-peak segment $s_1$, the transition segment $s_2$, and the highest-$k$ segment $s_3$.}
  \label{fig:piecewise}
\end{figure*}

To quantify this departure of self-similarity in all simulations, we fit the spectra to the right of the peak and beyond the particle scale at $kd_0=1$. For each selected time, we fit the post-peak spectrum in log-log space using one, two, and three piecewise-linear segments. The preferred model is selected by the Bayesian information criterion. For every sampled time in each case, the three-segment model is preferred, indicating that the post-peak spectrum is not a single power law. We denote the three fitted slopes by $s_1$, $s_2$, and $s_3$, ordered from lower to higher $k$. Figure~\ref{fig:piecewise} shows the log-time evolution of the three fitted indices.

The fitted indices are not constant. The lower-$k$ post-peak segment $s_1$ is relatively shallow and changes modestly over time, while the transition and highest-$k$ indices can evolve substantially. The highest-$k$ segment $s_3$ has values close to $-4$ to $-5$, comparable to kinetic-transition-range steepening reported in space-plasma observations and kinetic turbulence simulations \citep{sahraoui2009,sahraoui2010,voitenko2011,zhdankin2017,zhdankin2018}.
While $s_1$ has reached saturation at $t/l_0>50$,
$s_2$ and $s_3$ still increase at the end of the simulations.
Nevertheless, it is clear that the post-peak spectrum contains a resolved time-dependent structure in our simulations, indicating broken self-similarity.

To ask whether the three-segment model is tied to a kinetic scale, we compare the fitted break locations with a representative Larmor radius. The Larmor radius used here is estimated from the particle-averaged Lorentz factor and the rms total magnetic field,
\begin{equation}
  \rho_L=\frac{\langle\gamma\rangle m c^2}{e b_{\rm rms,tot}},
  \qquad
  b_{\rm rms,tot}=\langle b_x^2+b_y^2+b_z^2\rangle^{1/2},
  \label{eq:larmor_definition}
\end{equation}
where $c=1$ in our units. This $\rho_L$ is therefore a global diagnostic kinetic scale based on rms quantities. 
The vertical dot-dashed lines seen in Figure~\ref{fig:full_spectrum} indicate the locations of $k\rho_L=1$ for the two spectra.
They are all located near the second break point.

Table~\ref{tab:piecewise} summarizes the median break positions and slopes at $t/l_0=19.9$ and $t/l_0=124.9$. Here, $k_{b1}$ and $k_{b2}$ denote the wavenumber positions of the first and second fitted spectral breaks, respectively. The first break lies near $k_{b1}\rho_L\sim0.5$--$0.9$, and the second one lies near $k_{b2}\rho_L\sim1.0$--$2.4$, indicating that Larmor radius scale may be related.

\begin{table*}
\centering
\caption{Median post-peak piecewise spectral-break wavenumbers and slopes. The quantities $k_{b1}$ and $k_{b2}$ are the positions of the first and second fitted spectral breaks. The fits use $kd_0\leq1$.}
\label{tab:piecewise}
\small
\begin{tabular}{lccccccc}
\toprule
Case & $k_{b1}\xi_b$ & $k_{b2}\xi_b$ & $k_{b1}\rho_L$ & $k_{b2}\rho_L$ & $s_1$ & $s_2$ & $s_3$ \\
\midrule
$\sigma_0=0.25$ & 3.98 & 7.97 & 0.48 & 0.96 & -1.90 & -2.77 & -4.52 \\
$\sigma_0=1$    & 7.46 & 13.92 & 0.58 & 1.09 & -1.79 & -3.00 & -4.69 \\
$\sigma_0=4$    & 8.37 & 16.80 & 0.64 & 1.28 & -1.72 & -2.90 & -4.50 \\
$\sigma_0=16$   & 7.13 & 16.57 & 0.71 & 1.60 & -1.62 & -2.49 & -4.24 \\
$\sigma_0=64$   & 5.78 & 14.94 & 0.93 & 2.39 & -1.80 & -2.82 & -4.12 \\
\bottomrule
\end{tabular}
\end{table*}

\section{Possible Origins of the Slow Decay with magnetization dependence}

The broken self-similarity implies a departure of the global evolution time scale from the local reconnection time. In an MHD model with resistivity $\eta$, the local reconnection time scale is determined by the Lundquist number $S$ and the Alv\'en timescale \citep{loureiro2005, loureiro2007},
\begin{align}
    \tau_{\rm loc}=S^{1/2}\frac{\xi_b}{v_A}=\left(\frac{\xi_b v_A}{\eta}\right)^{1/2}\frac{\xi_b}{v_A}.
\end{align}
For a self-similar system, the global evolution time scale is only subject to an additional proportional factor, thus one has $t\sim t_{\rm decay}\propto\tau_{\rm loc}\propto \xi_b^{3/2}v_A^{-1/2}$.
Assuming power-law evolution $\xi_b\sim t^q$ and $v_A\sim b\sim t^{-q/2}$, it provides the relation $3q/2+p/4=1$. Together with $p=2q$ given by the conservation law, one recovers the prediction $p=1$ and $q=1/2$.

A collisionless system does not have a well-defined resistivity $\eta$ generally. The local timescale is replaced by 
\begin{align}
    \tau_{\rm loc,PIC}=\frac{\xi_b}{\epsilon_{\rm rec}v_A},
\end{align}
where $\epsilon_{\rm rec}$ represents dimensionless reconnection rate. Even though the reconnection rate $\epsilon_{\rm rec}$ could still be statistically constant, without self-similarity, $t_{\rm decay}\propto \tau_{\rm loc,PIC}$ is no longer valid, and this relationship is likely to be time dependent. 

Moreover, the deviation from the MHD scaling is related to the modification of the Alf\'ven velocity $v_A={B}/{\sqrt{4\pi\rho}}$ in a relativistic collisionless system. One correction term comes from the fact that enthalpy inertia and magnetic field inertia must be counted, which modifies the denominator. Another correction is that pressure anisotropies change the effective tension of magnetic field lines, thus the numerator must be modified. In general, the expression of Alf\'ven velocity is \citep{gedalin1993}
\begin{equation}
  v_{A}
  =
  \frac{B_{\rm tot}}
  {\sqrt{4\pi h+B_{\rm tot}^2}}
  \sqrt{{\cal T}} ,
  \label{eq:gedalin_va}
\end{equation}
where $h$ is the enthalpy density of the system, and
\begin{equation}
  {\cal T}=1+\frac{\beta_\parallel\Delta}{2},
  \label{eq:tension_factor}
\end{equation}
with
\begin{equation}
  \Delta=\frac{p_\perp}{p_\parallel}-1,
  \qquad
  \beta_\parallel=\frac{8\pi p_\parallel}{B_{\rm tot}^2}.
  \label{eq:delta_def}
\end{equation}
In principle, the Alf\'ven velocity must be calculated in the fluid frame. We measured bulk fluid velocities in our system and found that even for the highest initial background magnetization $\sigma_0=64$, the fluid elements remain well sub-relativistic (with $\gamma_{\rm max}\sim1.2 $ and a strong power law decay in the number distribution) in our main dynamical window, $t/l_0 > 20$.
The frame difference is therefore ignored in this case.

With the 2D nature of our system, reconnection happens in the $x-y$ plane,
and the effective velocity in Equation~\eqref{eq:gedalin_va} is projected into the $x-y$ plane.
Thus, the effective Alf\'ven velocity for our system is
\begin{align}
    v_{A, \rm eff}=\frac{b}{\sqrt{4\pi h+b^2+b_\text{g}^2}}\sqrt{\cal{T}}.\label{va_eff}
\end{align}

\begin{figure}
  \centering
  \includegraphics[width=1.\linewidth]{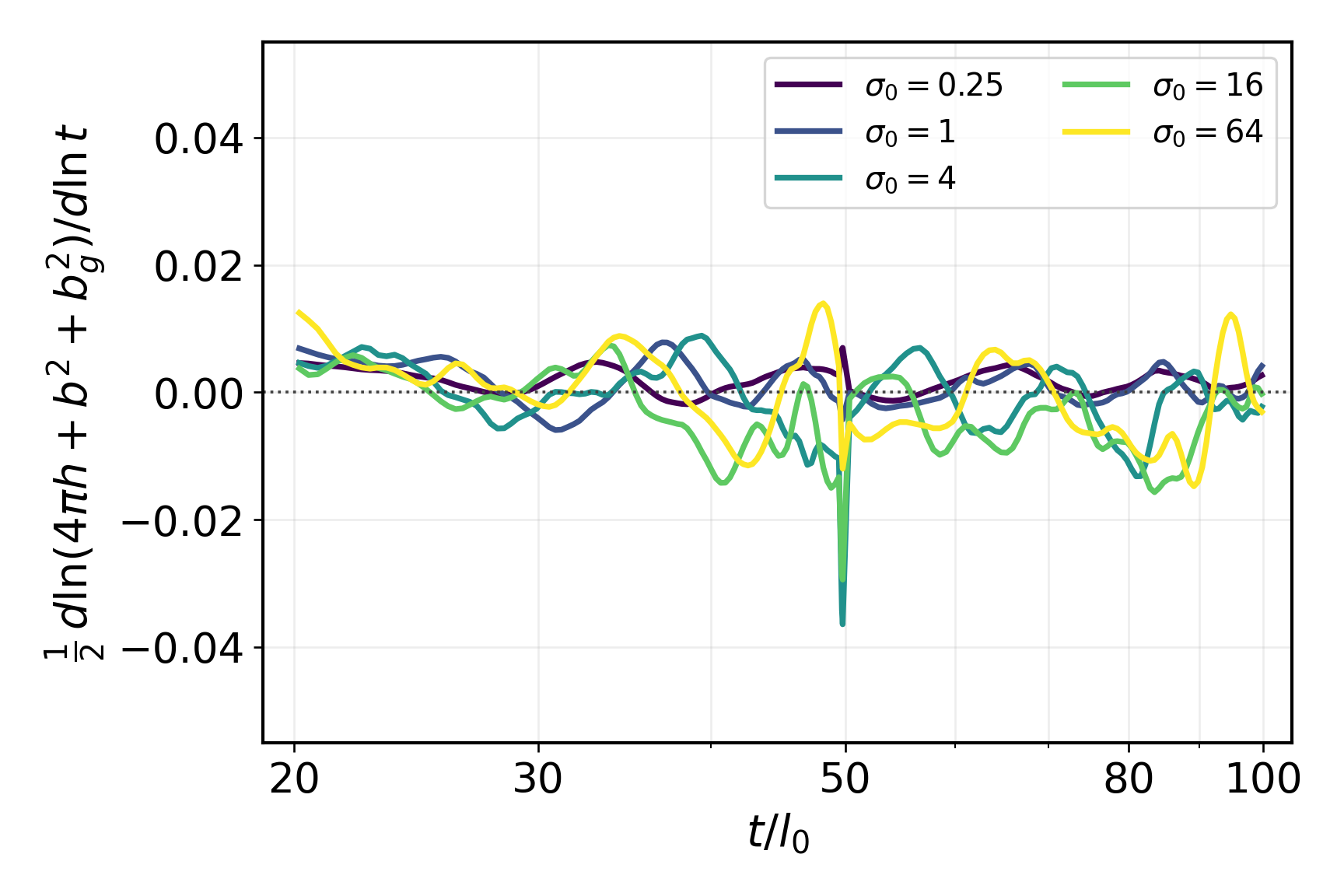}
  \caption{Logarithmic contribution of the denominator $4\pi h+b^2+b_\text{g}^2$ in Equation~(\ref{va_eff}) to the Alfv\'en-speed scaling. The plotted quantity is $(1/2)d\ln(4\pi h+b^2+b_\text{g}^2)/d\ln t$ over the main dynamical interval.}
  \label{fig:alfven_denominator}
\end{figure}

The effect of pressure aniostropy is discussed in the following subsection. Here we focus on the denominator of Equation~\eqref{va_eff} and demonstrate that it is approximately constant in our system. In the double-adiabatic evolution picture \citep{gedalin1993}, enthalpy is expressed as $h=\rho+2p_\perp+p_\parallel/2$. With this expression, we calculate the rate of change of the denominator of Equation~\eqref{va_eff}. Figure~\ref{fig:alfven_denominator} shows that the logarithmic contribution of the denominator to the total power law decay remains close to zero, and thus it is not the dominant source of the slow decay.

The previous results and discussions have established an empirical puzzle: the large-scale spectrum and the energy--length relation retain an approximately power-law inverse-transfer structure, but the decay is slower than the MHD equal-island prediction. We now propose two possible origins of the slow decay. The first is an initial-population effect: our initialization does not create identical islands. A finite-width island ensemble can already produce a somewhat slower merger sequence \citep{zhou2021}. The second is a kinetic effect: pressure anisotropy can modify magnetic tension and introduce kinetic-scale magnetic structure \citep{kunz2014}, contributing to the magnetization dependency of local-to-global time-scale relation.

\subsection{Initial Island Population}

The initial magnetic field is constructed from a finite band of Fourier modes. Thus, it does not produce a population of identical circular islands. To connect the simulation setup with the island kinetic theory of \citet{zhou2021}, we identify magnetic islands in the initial flux function and measure their area $S$ and signed island flux $\psi=A_z(O)-A_z({\rm sep})$, where $O$ denotes the island O-point and ``sep'' denotes the relevant separatrix value. We normalize these quantities by the injection-scale estimates
\begin{equation}
  \hat S=\frac{S}{l_0^2},
  \qquad
  \hat\psi=\frac{\psi}{b_{\rm rms,0}l_0}.
\end{equation}

\begin{figure*}
  \centering
  \includegraphics[width=0.75\textwidth]{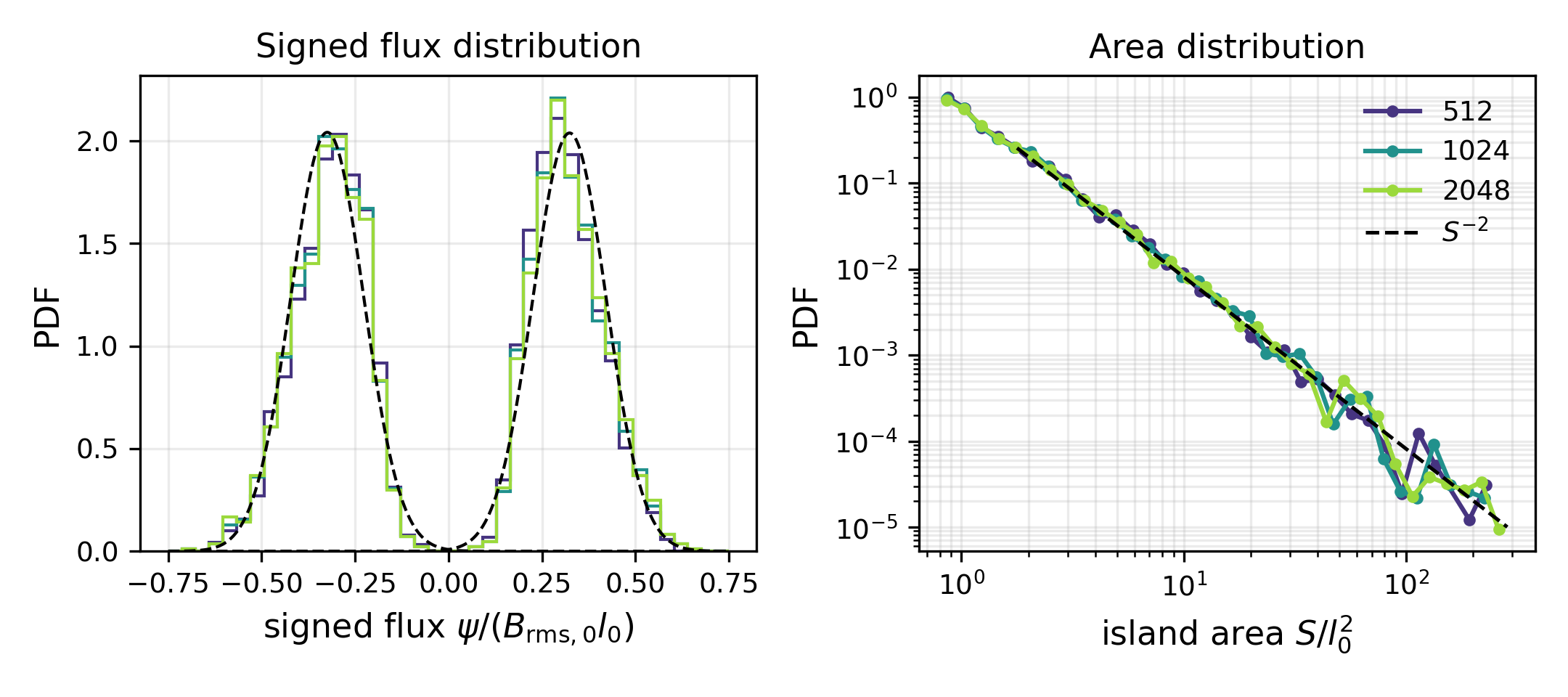}
  \caption{Initial magnetic-island distribution. Left: signed flux normalized by $B_{\rm rms,0}l_0$. The two polarity populations form two approximately Gaussian peaks, shown by the black dashed curves. Right: island area normalized by $l_0^2$, with a robust $S^{-2}$ tail.}
  \label{fig:initial_islands}
\end{figure*}

Figure~\ref{fig:initial_islands} shows that the signed flux distribution is bimodal. The positive-polarity and negative-polarity peaks are each approximately Gaussian, with nearly equal magnitude and width. The area distribution is much broader and has a tail close to
\begin{equation}
  P(S)\propto S^{-\alpha},\qquad \alpha\simeq2.05.
\end{equation}
This places the initial condition closer to a broad island ensemble than to a delta-function hierarchy\footnote{The island identification in Figure~\ref{fig:initial_islands} is performed on analysis grids up to $2048^2$, not on the full $8192^2$ PIC grid. This is sufficient for the initial-condition statistics. The highest initialized mode is $N_2=64$. The smallest initialized wavelength is $l_0=L/64$. A $2048^2$ analysis grid resolves this scale with $32$ cells per $l_0$, and the measured flux moments, median $\hat S$, and tail exponent are already stable between $1024^2$ and $2048^2$.}.

This measured distribution suggests one reason why the decay can be slower than the idealized equal-island hierarchy. \citet{zhou2021} found that non-trivial initial island distributions, including Gaussian and power-law distributions, still produce approximate power-law decay. However, their numerical island kinetic equation gives slightly shallower evolution than the delta-function hierarchy. The problem is a finite-width island population that contains islands with different areas and fluxes. The collision cross sections of two islands depend on the areas and fluxes. Thus, the merging process can not be described as a single series of hierarchical structures but multiple series with different time scales. In other words, since islands with similar sizes and fluxes are more likely to merge with each other, non-trivial island distribution results in a few different series of merger trees, thus results in a few different local timescales $\tau^i_{\rm loc, PIC}$. The global decay timescale is related to a few timescales with a non-trivial function $t_{\rm decay}(\tau^i_{\rm loc, PIC})$ and somehow makes it a slower power law decay.

This analysis identifies a common contribution to the slow decay though it does not measure how much of each fitted exponent is caused by the initial distribution. Moreover, it cannot by itself explain the systematic magnetization dependence because all five simulations start from the same normalized magnetic-field realization. The remaining trend with $\sigma_0$ must come from the other effects, not from the initial island population.

\subsection{Pressure Anisotropy and Kinetic-Scale Magnetic Structure}
\label{subsec:pressure_anisotropy}

All five runs start from the same normalized magnetic-field realization. The remaining magnetization dependence, therefore, likely comes from the kinetic plasma response during the decay. In the present diagnostics, pressure anisotropy is the best-supported candidate because it changes the magnetic tension that sets the Alfv\'enic part of the local merger time.

The origin of this anisotropy can be understood in the double-adiabatic picture \citep{cgl1956}. In the absence of rapid pitch-angle scattering, particles approximately conserve their magnetic moment, and the parallel motion has a longitudinal adiabatic invariant. In the form by \citet{cgl1956}, these give
\begin{equation}
  \frac{p_\perp}{nB_{\rm tot}}\simeq{\rm const},
  \qquad
  \frac{p_\parallel B_{\rm tot}^2}{n^3}\simeq{\rm const}.
  \label{eq:adiabatic_invariants}
\end{equation}
where $B_{\rm tot}=|\boldsymbol{B}|$. Equivalently, one may view the underlying single-particle invariants as $\mu\sim mv_{\perp,\rm th}^2/(2B_{\rm tot})$ and $J\sim\oint mv_{\parallel,\rm th}dl$. If density variations are not the leading effect, Equation~(\ref{eq:adiabatic_invariants}) reduces to $p_\perp/B_{\rm tot}\simeq{\rm const}$ and $p_\parallel B_{\rm tot}^2\simeq{\rm const}$. When island coalescence reduces the characteristic magnetic field, $p_\perp$ tends to decrease while $p_\parallel$ tends to increase. The plasma is then driven toward $p_\perp<p_\parallel$, i.e., $\Delta<0$, which reduces the tension factor ${\cal T}$ defined in Equation~(\ref{eq:tension_factor}).

\begin{figure*}
  \centering
  \includegraphics[width=0.75\textwidth]{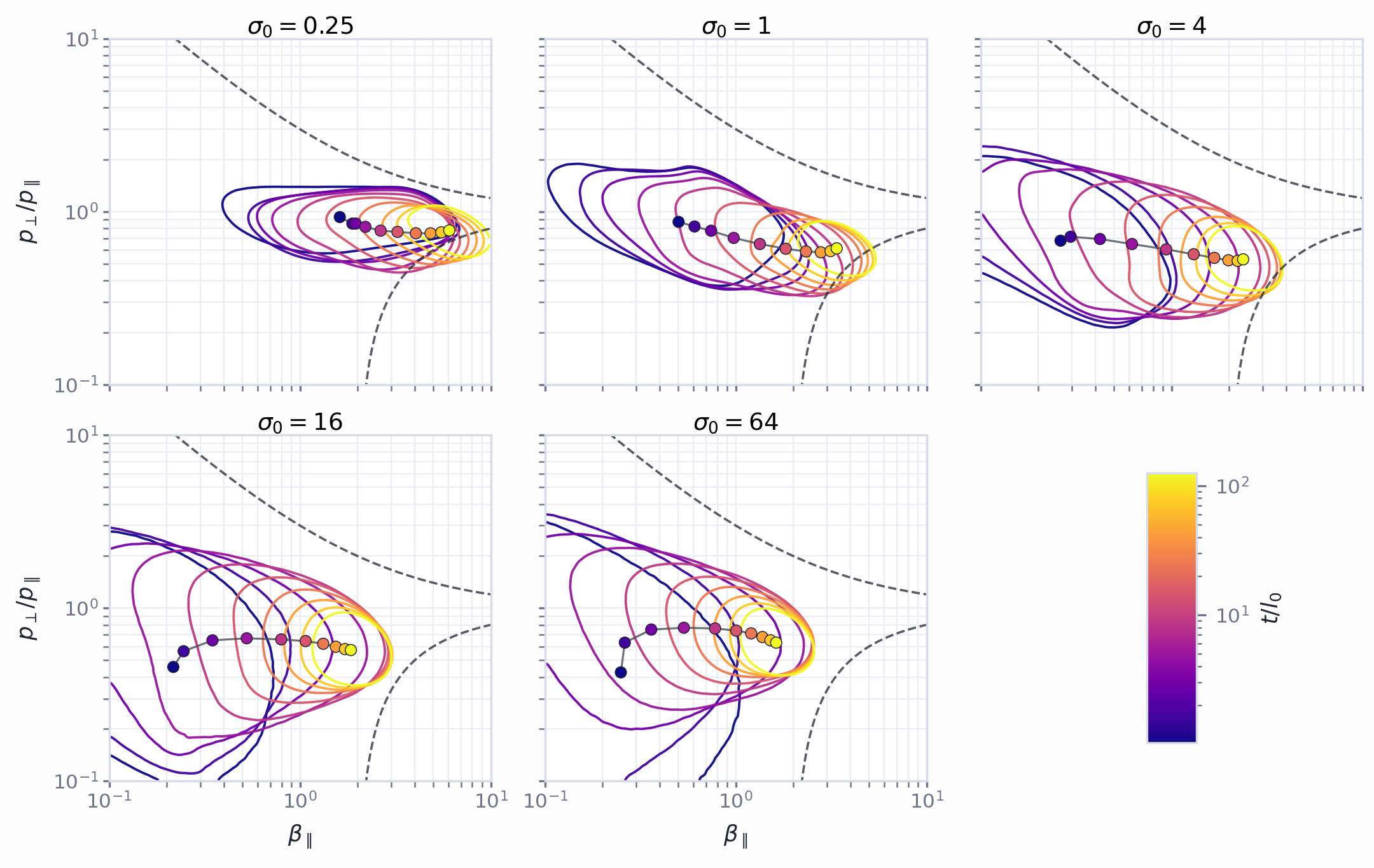}
  \caption{Pressure-anisotropy phase plots for the five magnetizations. Each panel shows the domain-averaged trajectory in $(\beta_\parallel,p_\perp/p_\parallel)$, colored by time. Each contour encloses $90\%$ of the plasma-cell probability distribution at one sampled time, with the outer $10\%$ lying outside the boundary. The dashed curves mark $\Delta\beta_\parallel/2=\pm1$, where $\Delta=p_\perp/p_\parallel-1$; the lower branch corresponds to a strong reduction of the effective magnetic tension.}
  \label{fig:anisotropy}
\end{figure*}

With this notation, we write
\begin{equation}
  v_{A,\rm eff}^2={\cal T}v_{A,0}^2,
  \qquad
  v_{A,0}^2=
  \frac{b^2}
  {4\pi h+b^2+b_\text{g}^2}.
  \label{eq:tension_va_relation}
\end{equation}
Figure~\ref{fig:anisotropy} shows that the simulations develop substantial pressure anisotropy in the main decay stage. Each panel plots the trajectory of the domain-averaged pressure state in the $(\beta_\parallel,p_\perp/p_\parallel)$ plane for one magnetization. Here, the color indicates a different time. The solid contour at each sampled time encloses the region containing $90\%$ of the plasma-cell probability distribution in the same variables. The dashed curves mark the schematic conditions $\Delta\beta_\parallel/2=\pm1$ (the tension factor ${\cal T}=1+\Delta\beta_\parallel/2$ equals 2 or 0) corresponding to mirror and fire-hose instability criteria, respectively. It clearly shows that all five cases approach the line of fire-hose instability. The $\sigma_0=0.25$ and $1.0$ cases have been caught by fire-hose instability, and the distributions move along the fire-hose instability criteria line at later times. It also shows that a bigger initial $\sigma_0$ gives weaker pressure anisotropy. This is not a surprise since pressure is less important in higher magnetization environment.

\begin{figure}
  \centering
  \includegraphics[width=1.0\linewidth]{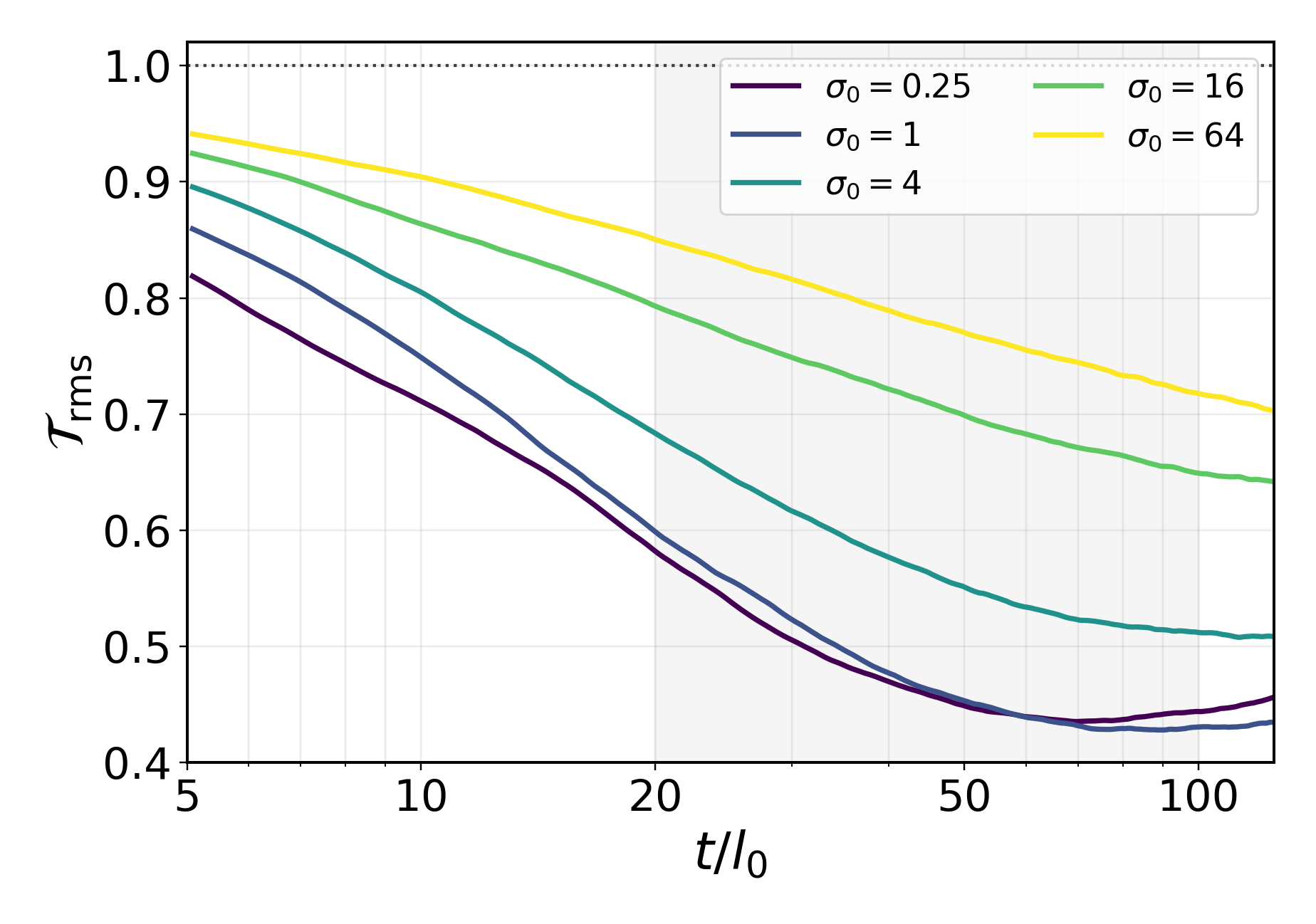}
  \caption{Log-time evolution of the rms effective-tension diagnostic ${\cal T}_{\rm rms}=(v_{A,\rm eff,rms}/v_{A,0,\rm rms})^2$. The shaded region marks the main dynamical interval $20\leq t/l_0\leq100$, and the dotted line marks ${\cal T}_{\rm rms}=1$. Lower magnetization runs show a stronger reduction of the effective tension.}
  \label{fig:teff_logtime}
\end{figure}

For a compact time-dependent diagnostic, we use the squared ratio of pressure-corrected and uncorrected rms Alfv\'en-speed estimates,
\begin{equation}
  {\cal T}_{\rm rms}
  =
  \left(\frac{v_{A,\rm eff,rms}}{v_{A,0,\rm rms}}\right)^2,
  \label{eq:teff_diagnostic}
\end{equation}
computed from the whole-domain rms outputs. Figure~\ref{fig:teff_logtime} shows its evolution in logarithmic time. The reduction of ${\cal T}_{\rm rms}$ is strongest at lower magnetization, consistent with the idea that pressure-anisotropy-modified tension contributes to the $\sigma_0$ dependence of the global decay time scale.

The second role of pressure anisotropy is the spectra. Anisotropy-regulated collisionless dynamics can generate or enhance magnetic structure near kinetic scales, rather than merely changing a large-scale Alfv\'en speed. This scale connection is well established in pressure-anisotropy-driven instability studies: firehose fluctuations are generated at ion-Larmor scales, mirror structures can grow to scales larger than the Larmor radius, and both can feed sub-Larmor or microscale magnetic fluctuations that regulate the anisotropy and feed back on the macroscopic field evolution \citep{schekochihin2008,kunz2014,melville2016,squire2017}. \citet{zhou2025} found a closely related spectral signature in collisionless inverse transfer: pressure-anisotropy-driven activity produces post-peak spectral plateaus or flattening, interpreted as small-scale magnetic-energy injection near the Larmor scale. This connects naturally to the spectra discussed in Section~3. Our post-peak magnetic energy spectrum contains two breaks; the higher-$k$ segment approaches the familiar steep kinetic range, while the lower-$k$ break may mark the onset of an anisotropy-driven flattened component rather than the ordinary sub-Larmor cascade alone. Pressure anisotropy thus offers a common physical route for two empirical facts: the local-to-global time-scale relation gives slower decay than the MHD prediction, and the post-peak spectrum is not strictly self-similar.

\section{Astrophysical Implications}

For astrophysical applications, the closest connection is cosmic magnetogenesis. Small-scale magnetic fields generated by kinetic processes must grow to larger coherence lengths before they can affect intergalactic or cluster-scale observables. MHD studies of decaying magnetic turbulence provide one route for this growth \citep{banerjee2004,brandenburg2015,brandenburg2017,durrer2013,subramanian2016}. \cite{zhou2025} showed that pressure-anisotropy-driven firehose activity can suppress inverse magnetic-energy transfer in marginally magnetized collisionless plasmas. Thus, Weibel instability-generated seed fields may fail to merge efficiently and may remain near kinetic scales. Our simulations further present another difficulty of inverse transfer: pressure anisotropy and Larmor-scale magnetic structure further slow down the growth of $\xi_b$, especially at low magnetization. This may strongly affect estimations of evolution histories based on current magnetic field strength and coherence length.

This point also connects to relativistic high-energy sources. \citet{zrake2014} provides an MHD benchmark: freely decaying, nonhelical relativistic turbulence shows inverse transfer, but the magnetic coherence length grows only as $t^{2/5}$. That growth was too slow to explain the optical polarization of gamma-ray burst afterglows if the field starts from microphysical plasma scales. The collisionless version of the same problem appears behind relativistic shocks, where Weibel-generated micro-turbulence decays downstream and affects synchrotron emission in gamma-ray burst and supernova-remnant blast waves \citep{medvedev1999,lemoine2012}. Our results address the kinetic side of this coherence-growth problem: pressure anisotropy and Larmor-scale structure can slow the transfer from plasma scales to larger magnetic scales.

The solar wind gives a nearby comparison case because it combines decaying or expanding turbulence, ion-scale spectral breaks, and pressure-anisotropy-regulated instabilities in a weakly collisional plasma \citep{bruno2005,sahraoui2009,sahraoui2010,hellinger2015,hellinger2017}. These are the same ingredients that appear in our diagnostics: an energy--scale relation at large scales, kinetic-scale spectral structure, and a time-dependent link between local activity and global evolution.

Other low-collisionality systems raise similar questions when the turbulence is no longer continuously driven. After galaxy-cluster mergers, intra-cluster turbulence can decay while dissipating through collisionless damping and particle interactions \citep{brunetti2007}. In such systems, kinetic plasma physics can change how magnetic energy, stress, and coherence length evolve.

\section{Conclusions}

We have shown that collisionless inverse transfer retains the energy--scale relation of the MHD island-merger picture but not its decay-time scaling. Across the five different magnetizations ($\sigma_0=0.25,1,4,16,64$) studied here, the simulations approximately satisfy $B^2\xi_b^2\simeq{\rm const}$. With power law evolution, $B^2\sim t^{-p}$ and $\xi_b \sim t^q$, we get $p=0.58,0.66,0.73,0.77,0.75$ and $q=0.27,0.31,0.33,0.38,0.39$ respectively. The spectral peak also moves to lower wavenumber ($k_{\rm peak} \sim t^{-q_{\rm peak}}$) with $q_{\rm peak}=0.43,0.43,0.48,0.56,0.50$. The self-similar evolution holds near MHD scale ($k_{\rm peak}^{-1}$) while breaks near kinetic scale ($k\rho_L \sim 1$).

The time dependence is nevertheless slower than the equal-island MHD prediction. We found $p<1$ and $q<1/2$ in all cases, but the exponents are not universal constants: both $p$ and $q$ vary systematically with $\sigma_0$. The spectrum gives the same message: $k_{\rm peak}^{-1}$ grows faster than $\xi_b$, and the post-peak spectrum contains two breaks near the Larmor scale, while it also flattens with time. The system is therefore not governed by a single self-similar scale.

The slow decay is likely contributed by two mechanisms. The initial condition produces a broad island-area distribution, which can introduce a non-trivial relation between local timescale and global decay timescale. The remaining magnetization dependence has a kinetic origin: pressure anisotropy is correlated with reduced effective in-plane magnetic tension and Larmor-scale spectral structure. A complete theory requires a local-to-global kinetic closure, based on reconnection-region diagnostics and time-dependent island statistics.

The present model is deliberately simple. By using a two-dimensional pair plasma with a prescribed guide field and a controlled initial spectrum, we can isolate the clean physics of the inverse-transfer energy--scale relation and its kinetic time-scale modification without mixing it with expansion, driving, radiative cooling, source geometry, or fully three-dimensional cascade effects. This controlled setting also defines the next steps. Three-dimensional simulations will be needed to test whether the same separation between the energy--scale relation and decay-time scaling survives when islands are replaced by flux ropes and when additional cascade channels are available. Electron--ion calculations can determine how mass asymmetry, Hall physics, and ion-scale pressure anisotropy modify the pair-plasma decay-time scaling. More detailed diagnostics should also identify reconnecting regions directly, compare volume-weighted and $E_z'$-weighted anisotropy measures, and follow time-dependent island or flux-rope statistics. In this sense, the present results are not a closed astrophysical model; they provide a clean kinetic baseline for future studies of collisionless inverse transfer in more realistic environments.

\section*{Acknowledgments}
This research is supported by the National
Key R\&D Program of China (grant No. 2023YFE0101200),
the National Natural Science Foundation of China (grant Nos.
12273022 and 12511540053), and the Shanghai Municipality
Orientation Program of Basic Research for International
Scientists (grant No. 22JC1410600).
YC acknowlegdes support from Shanghai Super Postdoc Program. HZ acknowledges support from the National Natural Science Foundation of China (No. 12403020), and the Qimeng project at Shanghai Polytechnic University. The authors also thank SuperComputing Network (SCNet) for the support during the code development.

\bibliographystyle{aasjournalv7.1}
\bibliography{references}

\end{document}